\documentclass[letterpaper, 10 pt, conference]{ieeeconf}  

\IEEEoverridecommandlockouts                              

\usepackage{graphicx} 
\usepackage{amsmath} 

\title{\LARGE \bf
Autofluorescence Bronchoscopy Video Analysis \\for Lesion Frame Detection*
}

\author{Qi Chang$^1$, Rebecca Bascom$^2$, Jennifer Toth$^{2}$, Danish Ahmad$^{2}$, and William E. Higgins$^{1}$
\thanks{*Research partially supported by NIH grant R01-CA151433.}
\thanks{$^{1}$Q. Chang and W. Higgins are with the School of Electrical
Engineering and Computer Science, Penn  State University, University Park, PA 16802, USA.
Email (W. Higgins): weh2@psu.edu}
\thanks{$^{2}$R. Bascom, J. Toth, and D. Ahmad are with the Dept. of
Medicine, Penn State Hershey Medical Center, Hershey, PA 17033, USA.}%
}

\begin{document}

\maketitle
\thispagestyle{empty}
\pagestyle{empty}

\begin{abstract}

Because of the significance of bronchial lesions as indicators of early lung cancer and squamous
cell carcinoma, a critical need exists for early detection of bronchial lesions.
Autofluorescence bronchoscopy (AFB) is a primary modality used for bronchial lesion
detection, as it shows high sensitivity to suspicious lesions.  The physician, however,
must interactively browse a long video stream to locate lesions, making the
search exceedingly tedious and error prone.   Unfortunately, limited research has
explored the use of automated AFB video analysis for efficient lesion detection.
We propose a robust automatic AFB analysis approach that distinguishes informative and
uninformative AFB video frames in a video.  In addition, for the informative frames,
we determine the frames containing potential lesions and delineate candidate lesion regions.
Our approach draws upon a combination of computer-based image analysis, machine learning,
and deep learning.  Thus, the analysis of an AFB video stream becomes more tractable.
Tests with patient AFB video indicate that $\ge$97\% of frames were correctly
labeled as informative or uninformative.  In addition, $\ge$97\% of lesion
frames were correctly identified, with false positive and false negative rates $\le$3\%.

\indent \textit{Clinical relevance}---The method makes AFB-based bronchial lesion
analysis more efficient, thereby helping to advance the goal of better early lung
cancer detection.
\end{abstract}

\vspace*{-8pt}

\section{INTRODUCTION}

\vspace*{-4pt}


Lung cancer remains the world's leading cause of cancer death \cite{bray2018global}.
Lung cancer begins when lesions develop in the bronchial epithelium of the lung mucosa
(thin walls of the airways).   Such bronchial lesions can eventually evolve
into squamous cell carcinoma and possibly indicate the growth of other more distant
lung tumors \cite{vanBoerdonk-AJRCC2015}.  This motivates a
new trend toward early detection of bronchial lesions and also partly drives the
search for biomarkers signaling lung cancer risk \cite{vanBoerdonk-AJRCC2015,
Billatos-CCR2018}.  Physicians use bronchoscopy to search the
airways for such lesions  \cite{Wisnivesky-Chest2013,Inage-CCM2018}.
A primary modality for lesion detection is autofluorescence bronchoscopy (AFB),
which transmits a light source that helps measure fluorescence and absorption
characteristics of airway wall tissue \cite{Inage-CCM2018}.  In particular, AFB
highlights differences between the normal epithelium and developing epithelial lesions.
Unlike standard white-light bronchoscopy, AFB exhibits high sensitivity to suspicious
bronchial lesions.  Thus, AFB figures prominently in several recent multi-center
early-detection studies \cite{Spiro-Thorax2016}.

Current usage of AFB forces the physician to search for lesions in a long, redundant
video stream and rely on interactive visual criteria to detect lesions
\cite{Wisnivesky-Chest2013,Inage-CCM2018,Byrnes-IEEEBME2019}.  Because a video stream contains
thousands of frames, the search is an impractical, highly time-consuming, error-prone task.
In particular, a typical video stream contains a high percentage of misleading uninformative
frames, obscured by water, mucus, and motion blur.  Other uninformative frames are
degraded by overexposure, as when the bronchoscope is too close to an airway wall.
While a few works have applied computer-based analysis to AFB video, they all
have severe limitations  \cite{bountris2009combined,
Zheng2016,finkvst2017classification,feng2018classification}:
1) don't distinguish informative and uninformative frames;
2) require manual interaction to identify a lesion frame; and
3) entail manual interaction to define a lesion.

We propose a automatic AFB analysis approach, drawing upon computer-based
image analysis, machine learning, and deep learning to detect informative
AFB video frames in a video stream.  Furthermore, for those frames deemed
informative, our approach determines which frames contain potential bronchial
lesions and delineates candidate lesion regions.  In this way, the analysis of
a long AFB video stream becomes more practical and helps focus subsequent
detailed classification.   Tests with human lung-cancer patient data demonstrate
the efficacy of our approach.

\vspace*{-8pt}

\section{METHODS}

\vspace*{-4pt}

Given a true-color three-channel RGB AFB video frame
${\bf I}$ (frame size = 720$\times$720 pixels (HD) or 360$\times$360 pixels (SD)),
our method has three stages: (1) Video
Preprocessing; (2) Frame Classification; (3) Lesion Analysis.
Video Preprocessing applies image-processing
methods to prepare the input for further analysis, while Frame Classification and
Lesion Analysis draw upon machine learning and deep learning.
The final outputs are:  (1) classification of ${\bf I}$ as
informative or uninformative; and (2) if ${\bf I}$ is classified as
informative, a decision on its likelihood of containing a lesion along
with candidate lesion locations.
\vspace*{-4pt}

\subsection{Video Preprocessing}

\vspace*{-4pt}

Preprocessing outputs three feature images used later by Frame Classification.
To do this, it first finds ${\bf I}$'s foreground/back\-ground regions and
isolates overexposed foreground regions, constituting useless uninformative
regions.  Focusing on the foreground reduces misclassifications arising in
mucosal areas beyond the bronchoscope's usable field of view.
To begin, ${\bf I}$ is downsampled to 180$\times$180 to reduce computation
and noise.  Also, since the spatial transitions between the informative foreground
and overexposed regions tend to be gradual, downsampling helps separate these
regions.  Next, the frame's intensity component ${\bf I}_{\rm I}$ and
green channel ${\bf I}_{\rm G}$ undergo separate computations.

Multi-level Otsu-based thresholding, which improves
upon Finkvst's AFB approach \cite{finkvst2017classification}, roughly
partitions ${\bf I}_{\rm I}$ into background, normal exposure, and overexposed regions.
Next, active contour analysis, which derives contours for
regions exhibiting weak edge gradients, combines the normal and overexposed
regions into a single foreground mask image ${\bf M}_{\rm fore}$
\cite{chan2001active}.  Also, gradient $\nabla {\bf I}_{\rm I}$ is computed.
(The Sobel operator is used for all gradient calculations.)

Because normal bronchial tissue fluoresces primarily in the green spectral
range (we use an Onco-LIFE AFB system in all studies), overexposure, which
predominates in normal regions, is more accurately defined via the green
channel ${\bf I}_{\rm G}$.  In particular, overexposed ${\bf I}_{\rm G}$ pixels
exhibit high intensity and low textural entropy.  The intensity feature at each
pixel is found by iterating a morphological bottom-hat transform that smooths
local irregularities in high-intensity neighborhoods:

\vspace*{-12pt}

\begin{equation}
	\begin{aligned}
	& {\bf I}_{\rm int} = {\bf I}_{\rm G}\\
	& {\rm For} \ i = 1, \dots 4\\
	& \hspace*{0.25in}  {\bf I}_{\rm int} = {\bf I}_{\rm int} + \left[ \left({\bf I}_{\rm int} \bullet {\bf B}_i\right) - {\bf I}_{\rm int} \right] \label{eq:bottom-hat}\\
	& {\rm EndFor}
	\end{aligned}
\end{equation}

\vspace*{-10pt}

\noindent where ``$\bullet$" signifies a morphological closing,
${\bf B}_i,\ i = 1,\dots , 4, $ are 1$\times$3 structuring elements within a
3$\times$3 neighborhood situated at 45$^o$ rotations (${\bf B}_1$ at 0$^o$,
${\bf B}_2$, at 45$^o$, etc.), and ${\bf I}_{\rm int}$ is the
final intensity feature image.  The entropy feature is computed at each pixel
$(x,y)$ using

\vspace*{-12pt}

\begin{equation}
{\bf I}_{\rm en}(x,y) = 1 + \frac{1}{\log_2(25)}\sum_{p \in {\bf P}} p\cdot\log_2(p)
\label{eq:entropy}
\end{equation}

\vspace*{-10pt}

To compute (\ref{eq:entropy}), we first calculate $|\nabla {\bf I}_{\rm G}(x,y)|$
normalized to the range [0,1] over a 5$\times$5 pixel neighborhood about $(x,y)$.
These values are then used to estimate a discrete pdf ${\bf P}(\cdot)$; only
non-zero values of $p \in {\bf P}$ are summed in (\ref{eq:entropy}).
Quantity ${\bf I}_{\rm en}$, which gives high
values for low-entropy pixels, then undergoes Gaussian smoothing ($\sigma = 5$,
mask size = 21$\times$21) to filter irregularities.  Finally, a
trained support vector machine (SVM) uses ${\bf I}_{\rm int}$ and
${\bf I}_{\rm en}$ to identify overexposed pixels.
(Section \ref{subsec:Train} discusses  method training.)  This gives
overexposure mask image ${\bf M}_{\rm over}$.

\vspace*{-12pt}

\subsection{Frame Classification}

\vspace*{-6pt}

Frame Classification uses the raw feature images
$\nabla {\bf I}_{\rm I}$,  ${\bf M}_{\rm fore}$, and ${\bf M}_{\rm over}$ \, to
flag uninformative frames.  Uninformative frames are characterized by a lack of
textural information and/or an excess of overexposed pixels.  To begin,

\vspace*{-4pt}

\begin{equation}
{\bf M}_{\rm inf} = {\bf M}_{\rm fore} - {\bf M}_{\rm over}
\end{equation}

\vspace*{-4pt}

\noindent identifies the informative foreground pixels in ${\bf I}$.
Next, seven features, partly inspired by previous white-light endoscopy
research, are computed \cite{Oh-MIA2007,Mctaggart-SPIEMI2019}:

\vspace*{-20pt}

\begin{eqnarray*}
	\begin{aligned}
	|\nabla| \mbox{ mean }\ \alpha &= \frac{1}{N}\sum |\nabla {\bf I}_{\rm I}|\\
	|\nabla| \mbox{ variance }\ \beta &= \frac{1}{N}\sum \left(|\nabla {\bf I}_{\rm I}| - \alpha \right)^2
\end{aligned}
\end{eqnarray*}

\vspace*{-2pt}

\begin{eqnarray}
	\begin{aligned}
	\mbox{Canny edge pixels}\ \gamma &= \frac{1}{N}\sum \Omega\left(\nabla {\bf I}_{\rm I}\right)\\
	\mbox{darkness }\ \rho &= \frac{1}{N}\sum {\bf I}_{\rm I}\\
	\mbox{sharpness }\ \ \epsilon &= \frac{1}{N}\sum \Phi({\bf I}_{\rm I})\\
	\mbox{sharpness variance}\ \ \zeta &= \frac{1}{N}\sum \left( \Phi({\bf I}_{\rm I}) - \epsilon \right)^2\\
	\mbox{overexposure ratio }\ \ \eta &= {\rm card}\left({\bf M}_{\rm over} \right)/
	{\rm card}\left({\bf M}_{\rm fore} \right)
	\end{aligned}
\label{eq:features}
\end{eqnarray}
All sums in (\ref{eq:features}) are computed over pixels constituting
${\bf M}_{\rm inf}$.  Quantity
$N = {\rm card}\left({\bf M}_{\rm inf}\right)$ denotes the total
number of informative pixels, where ${\rm card}\left( \cdot \right)$ is
set cardinality.  Function $\Omega(\cdot)$ is the Canny edge detector, based on
the gradient and nonmaxima suppression, to identify strong edge pixels.  Also,
for each $(x,y)$, $\Phi(x,y)$ finds the maximum difference between
${\bf I}_{\rm I}(x,y)$ and one of its 8-neighbors:

\vspace*{-8pt}

\begin{equation}
	\displaystyle \Phi(x,y) = \max_{(a,b) \in N_8(x,y)}\, |{\bf I}_{\rm I}(x,y) - {\bf I}_{\rm I}(a,b)|
\end{equation}

\vspace*{-4pt}

Finally, a pre-trained boosted decision tree uses the 7 features of (\ref{eq:features})
to arrive at frame decisions.  As alternatives, we also consider a two-layer
shallow neural network, which uses the features (\ref{eq:features}), and the ResNet-101
convolutional neural network (CNN), which uses the entire image instead of features
\cite{He-CVPR2016}.  Training details appear in Section \ref{subsec:Train}.

\vspace*{-4pt}

\subsection{Lesion Analysis}

\vspace*{-4pt}

An informative frame must show evidence of containing lesions for it to
be flagged as containing lesions.  Early AFB research relied on a simple
Red-to-Green ratio threshold test $\frac{\rm R}{\rm G} > 0.53$
to distinguish lesion and
normal areas \cite{kusunoki2000early}, with Finkvst using
the R and G distributions over a frame \cite{finkvst2017classification}.
Unfortunately, variations in illumination, scope-to-wall distance, and
surface ripples make the method unsuitable.

We perform lesion analysis by applying two trained SVMs and seeded region growing.
Each SVM draws on full-resolution versions of ${\bf I}$, ${\bf M}_{\rm fore}$,
and ${\bf M}_{\rm over}$.  To begin, SVM \#1 detects high-confidence
lesion seeds, while SVM \#2 uses less stringent criteria to flag additional
potential lesion pixels.  To do this, the SVMs are trained to weigh a false
positive (FP) and false negative (FN) in each class (lesion, no lesion)
differently, where a positive decision means that a pixel belongs to a lesion.
In particular, SVM \#1 heavily penalizes false positives (FP misclassification
cost = 6, FN cost = 0.1), while SVM \#2 emphasizes limiting false negatives
(FP cost = 1, FN cost = 2).

Region growing next grows the lesion seeds into regions
by adding pixels contained in the potential-lesion pixel set
to give a final candidate lesion mask.  The lesion mask is then subdivided into
36$\times$36 boxes.  If any box contains $>50\%$ lesion pixels,
then the frame is deemed to contain a potential lesion.  Furthermore,
the percentage of boxes containing a lesion gives the likelihood of a
frame containing a lesion.

\vspace*{-4pt}

\section{RESULTS}

\vspace*{-4pt}

\subsection{Data Set and Method Training}
\label{subsec:Train}

\vspace*{-2pt}

The training and test sets were drawn from 39,899 AFB video frames
constituting four lung-cancer patient bronchoscopies performed at the Penn State
Hershey Medical Center and obtained under IRB-approved
protocols.  We took care to be unbiased in constructing
all training and testing sets.

\begin{figure}[ht]
	\hspace*{0.6in}\includegraphics[width =2.2in]{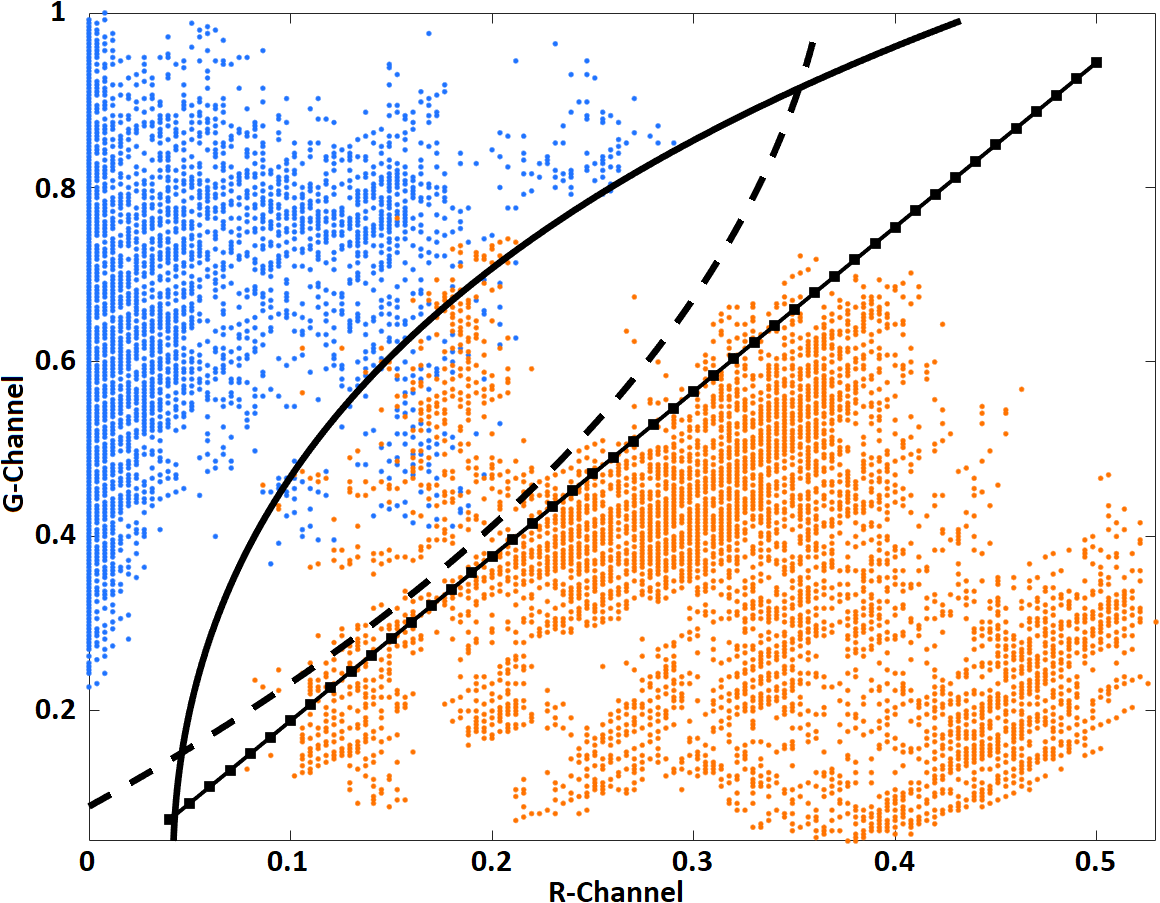}
    \vspace*{-3pt}
	\caption{Decision boundaries of SVM \#1 (dashed line), SVM \#2 (solid line) and
	$\frac{\rm R}{\rm G} = 0.53$ (solid line with dots) for the R(ed) and G(reen)
	channel training data.  Pixels under the SVM \#1 boundary are high-confidence
    lesion seeds, while pixels between the SVM \#1 and SVM \#2 boundaries are considered
	to be potential lesion pixels. Orange points denote preselected lesion
    pixels, and blue pixels are from normal tissue.}
	\label{fig:distribution}
\vspace*{-12pt}
\end{figure}

Briefly, we selected 25 frames to train the three SVMs used in our approach.  The
frames depicted a representative range of overexposed regions.  Also, 11 frames
contained distinct lesions, while 14 frames were judged normal. \\
\noindent {\it a.) Overexposure SVM training:}   Because a purely manually
derived ground-truth pixel set is difficult to define, we used the K-means
algorithm, which drew upon the ${\bf I}_{\rm int}$ and ${\bf I}_{\rm en}$
feature images, to form an initial segmentation of the 25 training frames
into high-average intensity (overexposure), mid-range intensity (informative foreground)
and dark areas (background).  A small erosion next produced more distinct normal
and overexposed regions.  From these 25 images, we selected 8000 pixels, half
from overexposed regions and half from normal exposure regions.   We next performed
Bayesian optimization via 5-fold cross-validation (90\% of data used for training;
10\%, testing) to achieve 98.2\% classification.\\
\noindent {\it b.) Training of two Lesion Analysis SVMs}:
220,000 pixels were manually selected from lesion and normal regions
in equal numbers (all 25 frames represented).  Each SVM was trained
using 5-fold cross-validation, with 200,000 pixels used to train and 20,000
pixels used to test.  The only difference between the SVMs were their respective
misclassification costs, as stated earlier.  Fig. \ref{fig:distribution}
depicts the derived decision boundaries.

\vspace*{-8pt}

To train the boosted decision tree, shallow neural network, and ResNet-101 
for in\-form\-ative-frame classification, the training set contained 1000 frames (500
informative, 500 uninformative).   Using random selection, 80\% of the frames in each
class were used to train and the remaining 20\% to test.  The boosted decision tree
was trained in a 5-fold cross-validation procedure. The model with lowest cross-validation
loss was used for classification tests, where the model consists of 11 distinct decision
trees (220 total decision operations) to boost the result.  In concordance with our boosted
decision tree's number of operations, we trained the shallow neural network using 220 hidden
neural nodes, with 20\% of the training data used to validate the network.
With the 7 features (\ref{eq:features}) serving as the input and ordered as
$\alpha, \beta, \ldots \eta$, we arrived at a trained network having
derived gains (2.74, 9.11, 2.13, 5.21, 4.73, 0.07, 1.93)
and offsets (0.03, 0, 0.06, 0.19, 0.007, 0, 0).
Finally, a pre-trained ResNet-101 was retrained as follows \cite{He-CVPR2016}:
1) freeze the beginning 10 layers to maintain the feature extractor;
2) replace the last fully connected layer with a new one having only two outputs;
3) use a SGDM (stochastic gradient descent with momentum) optimizer to train the network.
Overall, the boosted decision tree, shallow neural network, and ResNet-101 achieved
99\%, 90\%, and 99.5\% correct frame classification in the training data set.

\vspace*{-6pt}

\subsection{Experimental Results}

\vspace*{-4pt}

We tested our methods with the data below:\\
{\it Case 156} --- 4,683-frame HD AFB video having the following ground-truth
labels:  4,401 informative frames (93.9\% of the total) and 282
noninformative frames (6.1\%).  Within the informative-frame class, 2,394
were labeled as normal (51.1\%) and 2,007 as lesion (42.8\%).\\
{\it Case 157} --- 455 HD frames, with each frame randomly selected from a contiguous
10-frame segment of a 4,550-frame sequence.  Ground truth:  417, informative;
38, uninformative; all frames normal.\\
{\it Case 140} --- 400-frame SD video sequence consisting entirely of lesion frames.

\begin{figure}[ht]
	\includegraphics[width = 1.6in]{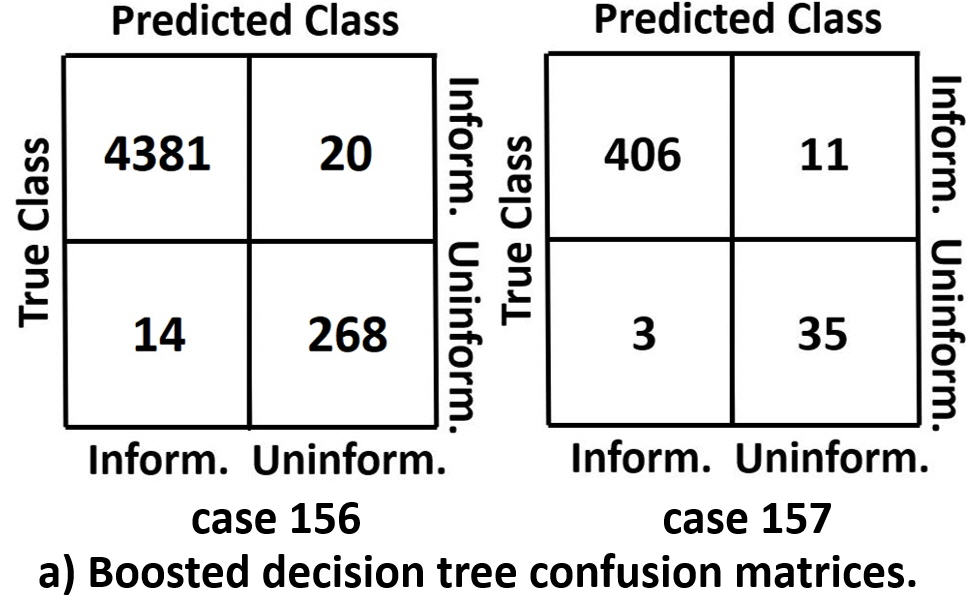}\ \
    \includegraphics[width = 1.6in]{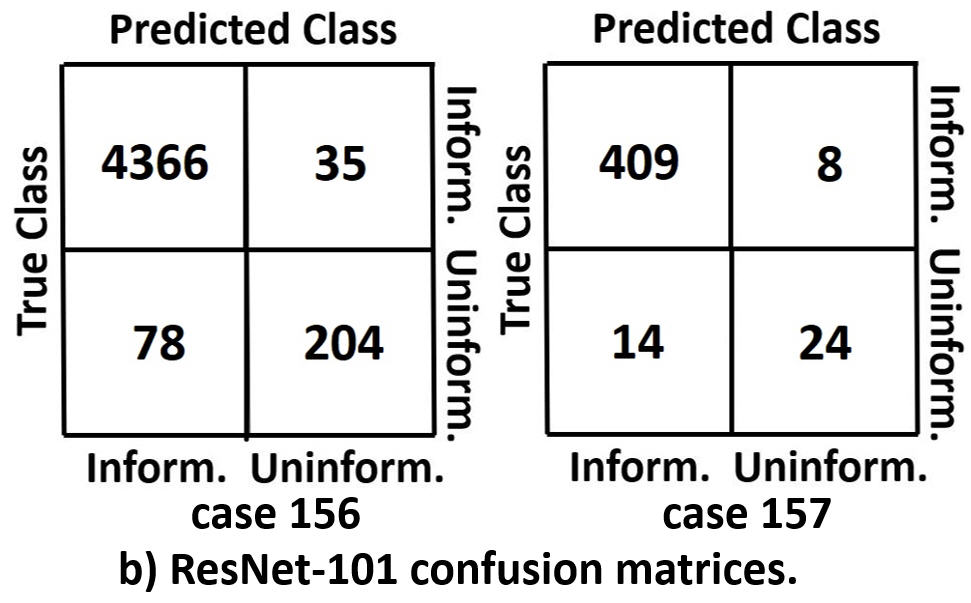}\\*[6pt]
    \includegraphics[width = \columnwidth]{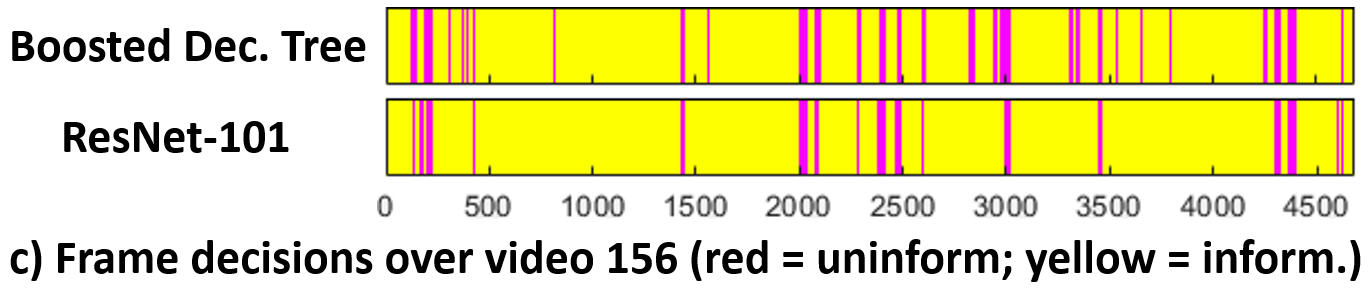}
    \vspace*{-16pt}
	\caption{Confusion matrix esults for frame classification.
(a) Boosted decision tree. (b) ResNet-101. (c) Method comparison of frame
decisions over entire case-156 4,683-frame sequence; yellow = informative
frame and red = uninformative frame.}
	\label{fig:156cm}
\end{figure}

Regarding Frame Classification,
Fig. \ref{fig:156cm}a-b show the 2$\times$2 confusion matrix results produced by the
boosted decision tree and ResNet-101.  Accuracy results for the methods were as
follows.  Boosted decision tree --- 99.3\% and 96.9\% for 156 and 157, respectively;
ResNet-101 --- 97.6\% and 95.2\%; shallow neural network --- 90\% and 89\%.
(Accuracy = (true positives + true negatives) / total frames.) Per the
nonnegative gains used for all features (\ref{eq:features}), the shallow neural
network supports the sufficiency of the features
used by the boosted decision tree.
Fig. \ref{fig:156cm}c shows how the boosted decision
tree more robustly detects uninformative frames than ResNet-101 (at least for this
sequence and our implementation).

\vspace*{-8pt}

\vspace*{-8pt}

Regarding Lesion Analysis tests, our approach gave the following results for
case 156.  For lesion frames, 97.1\% received correct labels, 2.7\% were
erroneously labeled normal, and 0.2\% were previously classified as uninformative.
For normal frames, 97.4\% received correct labels, 1.9\% received erroneous
lesion labels, and 0.7\% were previously classified as uninformative.
Next, for case 140, 100\% of the frames received correct lesion labels
(they all also were correctly identified as informative by the boosted decision tree).

\begin{figure}[htb]
	\begin{center}
		\begin{tabular}{cc}
			\includegraphics[width = 3.86cm]{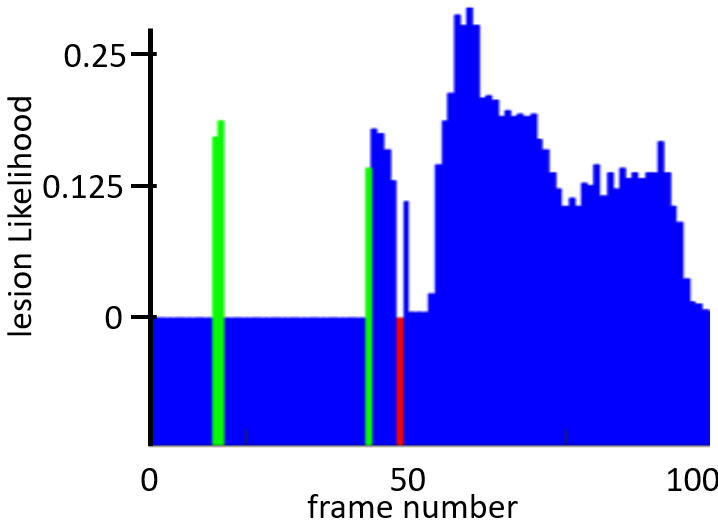} &\includegraphics[width = 2.85 cm]{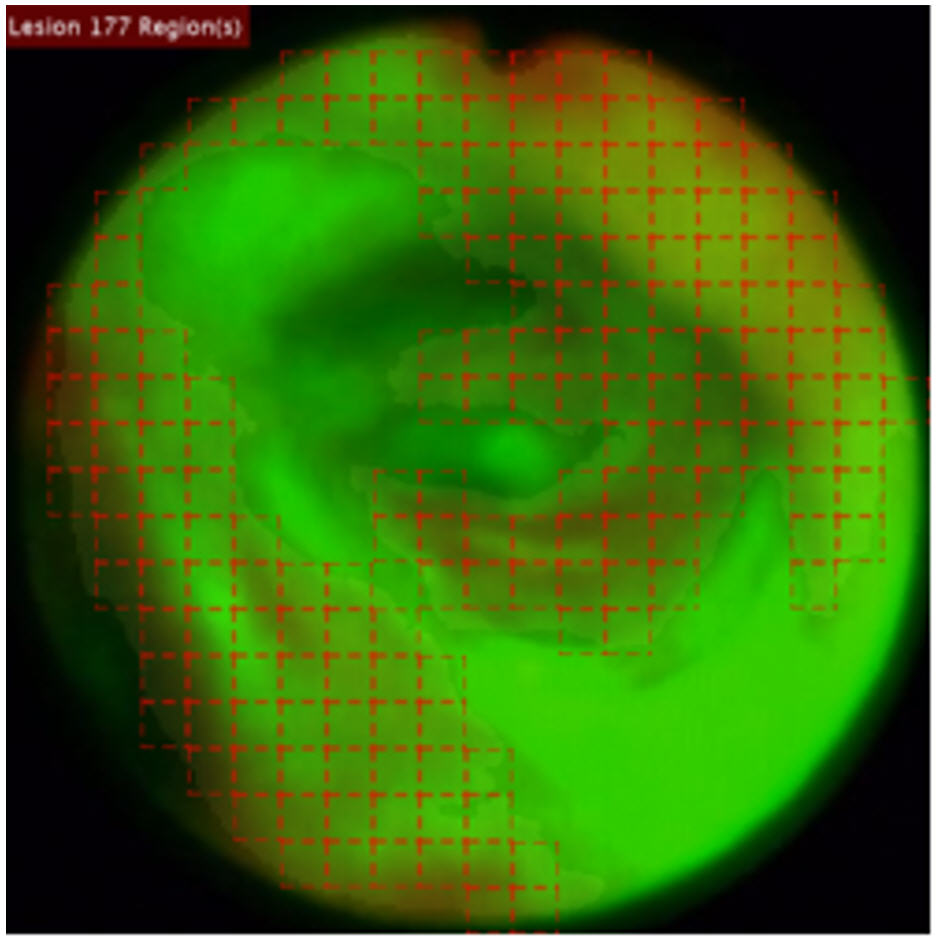}\\
			a) Lesion likelihood profile & b) Example lesion frame\\
		\end{tabular}
	\end{center}
    \vspace*{-10pt}
	\caption{Case 156 lesion analysis test. (a) Lesion likelihood profile over
		a 100-frame segment (blue = correct decision; green = false negative; red = false
		positive). (b) Flagged lesion frame (lesion likelihood = 0.44); red boxes denote locations of
		lesion tissue.}
	\label{fig:156-400frame}
\end{figure}

\begin{figure}[h]
	\hspace*{0.6in} \includegraphics[width = 5.95 cm]{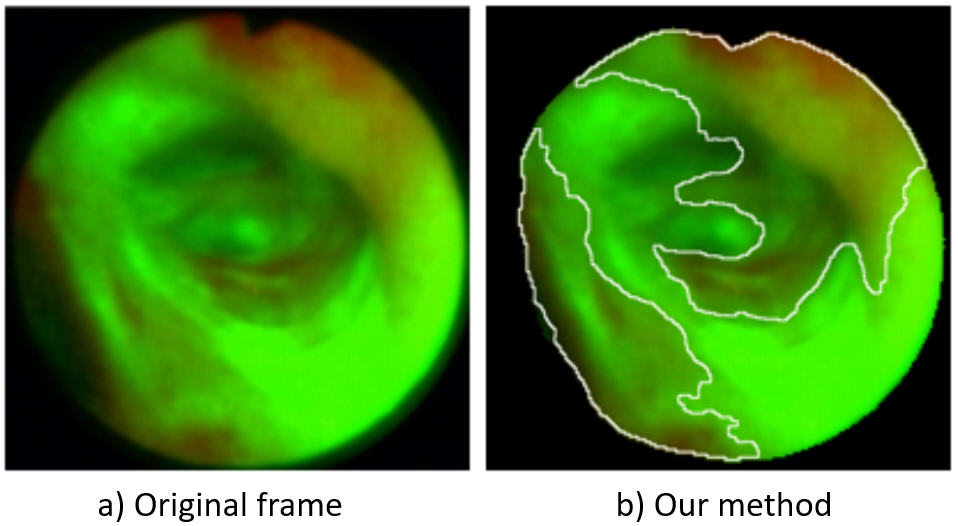}\\
	\hspace*{0.6in} \includegraphics[width = 5.95 cm]{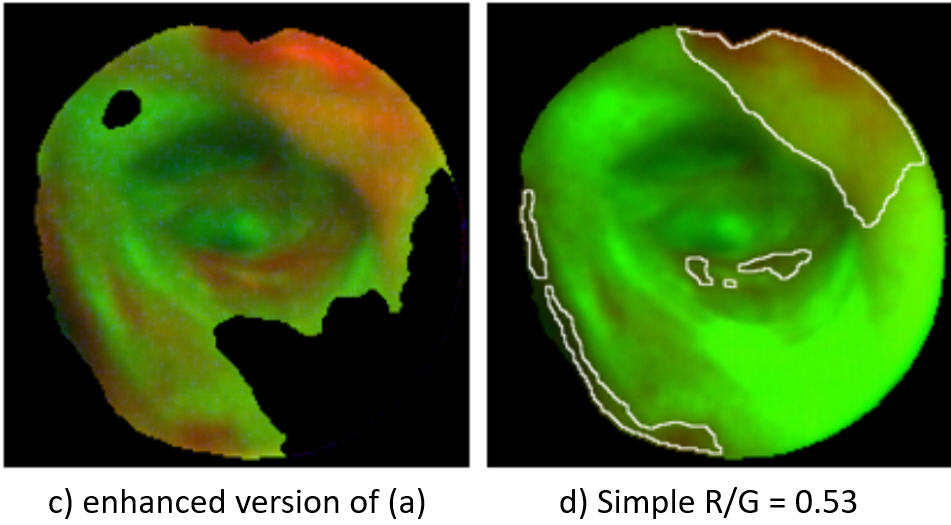}
    \vspace*{-10pt}
	\caption{Lesion segmentation example. (a) Original frame. (b) Our
		method --- suspect lesion areas delineated by the white contours. (c)
		Enhanced version of (a) \cite{bountris2009combined}. (d)
		$\frac{\rm R}{\rm G}$ ratio test.}
	\label{fig:lesions}
\end{figure}

As a third test, for a 400-frame sub-segment of case 156 containing only informative
frames, Fig. \ref{fig:acc} shows the superior performance of our method versus
two other methods:  the previously suggested Red-to-Green
$\frac{\rm R}{\rm G}$ ratio threshold test \cite{kusunoki2000early}
and another proposal using Fisher's linear discriminant.  A video
profile of our method's class assignments shows
that errors tend to be isolated frames (Fig. \ref{fig:156-400frame}a). \
Fig. \ref{fig:156-400frame}b illustrates a lesion-frame detection.
Finally, Fig. \ref{fig:lesions} shows the superiority of our lesion segmentation
method over the $\frac{\rm R}{\rm G}$ method.

\vspace*{-10pt}

\begin{figure}[htb]
	\centering
	\begin{tabular}{c|ccc}
		& Proposed & Fisher & R/G ratio\\
		\cline{2-4}
		Accuracy     & 98.8\%   & 94.3\% & 95.0\%\\
		sensitivity  & 99.6\%   & 92.0\% & 92.8\%\\
		specificity  & 97.1\%   & 98.5\% & 99.3\%
	\end{tabular}
	\caption{Lesion classification rates for case 156 (400-frame test).
		Fisher = Fisher's linear discriminant; R/G ratio = $\frac{\rm R}{\rm G}$ ratio test.}\label{fig:acc}
\end{figure}

\vspace*{-18pt}

\vspace*{-18pt}

\vspace*{-18pt}

\section{CONCLUSIONS}

We have proposed the first automatic analysis method for AFB video that:
1.) distinguishes uninformative frames from informative frames; and 2.)
determines which informative frames contain suspect bronchial lesions and
delineates lesion regions.  In this way, the search for
bronchial lesions becomes more efficient than interactive video browsing
and more accurate than simple $\frac{\rm R}{\rm G}$ thresholding.  Results show
a frame classification accuracy $\ge$97\% and a lesion detection accuracy
also $\ge$97\%.  Future work will focus on lesion classification and more extensive tests.

\noindent {\it Conflict of Interest:} Dr. Higgins and Penn State have financial interests in Broncus Medical, Inc.  These financial interests have been reviewed by the University's Institutional and Individual Conflict of Interest Committees and are currently being managed by the University and reported to the NIH.

\bibliographystyle{IEEEtran}
	\bibliography{Bibtex/ISBIrefs}

\end{document}